\documentclass[11pt,letterpaper]{article}
\pdfoutput=1

\usepackage[usenames,dvipsnames]{xcolor}
\usepackage{graphicx}
\usepackage{epstopdf} 
\usepackage{amsmath}
\usepackage{amssymb}
\usepackage{amsthm}
\usepackage{nicefrac}
\usepackage{url}
\usepackage{macros}
\usepackage{dblfloatfix}
\usepackage{MnSymbol}
\usepackage[normalem]{ulem}
\usepackage{subfigure}
\usepackage{wrapfig}
\usepackage[font=small,labelfont=bf]{caption}

\usepackage{natbib}
 \bibpunct[, ]{(}{)}{,}{a}{}{,}%
\usepackage[ruled]{algorithm2e}

\SetAlFnt{\small}
\SetAlCapFnt{\small}
\SetAlCapNameFnt{\small}
\SetAlCapHSkip{0pt}
\IncMargin{-\parindent}

\DeclareMathOperator*{\argmin}{arg\,min}
\DeclareMathOperator*{\argmax}{arg\,max}
\newcommand{\Expect}{\mathbb{E}}
\newcommand{\I}{\mathbb{I}}

\newtheorem{observation}{Observation}

\begin{document}


\title{Games and Meta-Games:\\ Pricing Rules for Combinatorial Mechanisms}
\author{Benjamin Lubin\\
Boston University School of Management\\
\texttt{blubin@bu.edu}\\
}
\date{First Version: January 2014\\
      This Version: March 20, 2015}
\maketitle

\begin{abstract}

\noindent In settings where full incentive-compatibility is not 
available, such as core-constraint combinatorial auctions and
budget-balanced combinatorial exchanges, we may wish to design
mechanisms that are as incentive-compatible as possible.  This paper
offers a new characterization of approximate incentive-compatibility
by casting the pricing problem as a meta-game between the center and
the participating agents.  Through a suitable set of simplifications,
we describe the equilibrium of this game as a variational problem.  We
use this to characterize the space of optimal prices, enabling
closed-form solutions in restricted cases, and numerically-determined
prices in the general case.  We offer theory motivating this approach,
and numerical experiments showing its application.

\end{abstract}


\section{Introduction}
%

Market mechanisms increase the welfare of their participants by
enabling the transfer of goods and services from those with low value
for their holdings to those with high value for those resources.  For
transacting large numbers of identical goods (like stocks), the double
auction formats used in financial markets work very well.  However,
there are many contexts in which goods are not uniform and
participants value different bundles of goods in complex ways, and we
thus need a more complex design.  This paper offers a novel way to
frame the problem of constructing mechanisms in these complex
settings, and then instantiates several concrete cases and solves
them.  The resulting mechanisms reduce the level of strategic
manipulation of bids in equilibrium, which decreases the cognitive
load of participants, in turn lowering the barrier to agent
participation and, even more importantly, raising the overall
efficiency of the outcome implemented.

\subsection{Combinatorial Mechanisms}

Combinatorial mechanisms are designed for settings where the
allocation of multiple goods to multiple participants is required, and
where the participants value packages of goods at either greater or
less than their constituant parts.  A prominent domain where such
complex mechanisms are needed is in markets for the transfer of
billions of dollars of radio spectrum rights from governments around
the world to mobile phone carriers \citep{kwerel2002proposal}.  In
such markets different frequency blocks have different properties and
thus different values; and a single frequency in different
geographical regions also can have wildly different values.  Bidders
are interested in packages across various geographic areas in order to
implement regional or national business plans.  Other such examples
include the transfer of aircraft landing rights at airports
\citep{ball2006auctions}, complex procurement and logistics markets
\citep{sandholm2007expressive}, and the lease of computational
resources in large-scale datacenters \citep{guevara2014market}.

Most of the mechanisms that have been designed for these settings are
\emph{Combinatorial Auctions (CA)}.  In CAs the goods are all offered
by a single seller, while multiple buyers use an expressive bidding
language to state their bid for different bundles of goods to the
mechanism.  The mechanism clears the market by determining the
welfare-maximizing allocation at the buyer reports, and then charges
these winning bidders according to a specific \emph{payment rule},
which will be a central concern of this paper.  If instead (1) there
is more than one seller; (2) these sellers are allowed to specify
complex reserve values over the goods they are offering; and (3)
participants can both buy and sell simultaneously -- then this
two-sided market is called a \emph{Combinatorial Exchange (CE)}.

Unlike the CAs that have been widely adopted, CEs have not been widely
used.  This is so, despite the need to reallocate goods in these same
settings in secondary markets that are thus by definition two-sided in
nature.  For example, the upcoming ``Incentive Auction'' planned by
the FCC is intended to facilitate reallocation of spectrum between
firms, each of which have combinatorial preferences
\citep{fcc2015incentive}.  The FCC plans to run a reverse auction to
recapture rights from existing holders followed by a forward CA to
offer these rights to new owners.  However, this design precludes
participants from swapping one set of goods for another without
potentially selling their goods and subsequently failing to obtain new
goods.  A full CE would solve this problem, but such designs have
rarely been proposed \citep{lubin2008ICE}.  One reason for this is a
lack of consensus about how best to price them.  The famed VCG
mechanism may run at a deficit for CEs, precluding its use without
undesirable subsidies; the Core may be empty for CEs, meaning that the
pricing rules that have recently gotten much attention in CAs (c.f. UK
bandwidth auctions \citep{Cramton2013SpectrumAuctionDesign}) are also
not available.  Our method provides high-quality pricing rules for
CEs, enabling their possible use in these billion dollar markets.

\subsection{The Problem With Payment Rules}
%

When designing a combinatorial mechanism, there are design challenges
in specifying the bidding language, a tractable winner-determination
algorithm, and effective activity rules.  But it is the payment rule
that \emph{directly} mediates the economic and game-theoretic aspects
of the mechanism.  The study of payment rules is thus of paramount
importance, and is our focus here. Specifically, there are a number of
desirable properties we would like our payment rules to have:
\begin{description}
\item[Individual Rationality (IR)] \emph{Ex-post}, individual bidders
  prefer to have participated in the mechanism, rather than remained
  outside it.  Concretely, this means bidders receive only weakly
  positive profits at their reported values.
\item[(Weak) Budget Balance (BB)] The total payments paid by bidders
  to the mechanism is weakly greater than the total disbursements by
  the mechanism; e.g. no subsidy is required to run the mechanism.
\item[Incentive Compatibility (IC)] In equilibrium, bidders should not
  have to strategize in submitting their bids; e.g. stating one's
  value truthfully should yield the maximum profit.
\end{description}
%
In addition to these properties of the payment rule, we are interested
in mechanisms that use the most natural winner determination rule, one
that implements the total value-maximizing allocation at the bids.  In
conjunction with incentive compatibility, this yields a mechanism
offering:
\begin{description}
\item[Efficiency (Eff)] The mechanism implements the total
  value-maximizing allocation at the participant's true values.
\end{description}
%


%
For CAs, only one mechanism satisfies all four of these properties:
the famous Vickrey-Clarke-Groves (VCG) Auction
\citep{nisan2007algorithmic}, which is a generalization of the
second-price or Vickrey \citep{Vickrey1961counterspeculation} auction
to settings with combinatorial preferences.  In VCG, a participant
pays his bid, less his marginal impact on the main economy (i.e., the
difference in social welfare between the main economy and the economy
where the participant is omitted).  However, as has often been pointed
out \citep{ausubel2006lovely}, VCG yields very low revenue to the
seller in the CA setting, which in turn creates incentives for
undesirable behavior on the part of participants -- for example, sybil
attacks, where a single firm bids under several identities.  Further,
VCG payments in a CA can be outside the Core \citep{shapley1974cores}.
This means that VCG prices may be sufficiently low that, \emph{ex-post}, the
seller may prefer to transact with a coalition of the bidders at a
price higher than VCG, rather than accept the VCG outcome.

To get around these problem, the most recent spectrum auctions have
been built around pricing rules that implement prices in the core, and
thus which may derive substantially more revenue than VCG
\citep{DayMilgrom2008coreAuctions}.  However, as soon as
prices are chosen to obey the Core constraints, we will have to relax
one of our aforementioned other properties.  In this work, we choose
to relax the incentive-compatibility property.


For CEs, the situation is even worse.  Even without core constraints,
the seminal Myerson-Satterthwaite Impossibility theorem
\citep{myerson1983efficient} tells us that no design can manifest our
desired four properties (IR, BB, IC, Eff) simultaneously in this
setting.\footnote{In CEs, the core is often empty, so it is not
  typically considered in defining payments.  However, it is possible
  to include it by instead targeting the minimal $\epsilon-$core (or
  nucleolus).}  Again, it is necessary to relax one of the four
conditions.  Because IR, BB and Eff are typically considered hard
constraints, we again choose to relax IC.

Even designs that are not explicitly combinatorial in nature can have
similar problems.  For example, the Generalized Second Price (GSP)
auction \citep{Edelman2007GSP} used by the major search engines to
sell the links in the sponsored section of their results pages is not
equivalent to VCG, and thus is not IC.  In this case, while VCG is
available, we are restricted from using it by a constraint on the
simplicity of the mechanism (e.g. single-dimensional prices).  One
instead might want a mechanism that is maximally IC, while still
obeying a constraint on simplicity.

In all these cases, what we want is a mechanism that from the
participant's perspective looks like VCG.  However, we want to charge
something slightly greater than VCG in order to meet the other other
requirements of the domain (e.g. to be in the core of a CA, attain
budget balance in a CE, or maintain simplicity in the sponsored search
setting).  Charging this extra amount will mean, of necessity, that
our mechanism will not be IC.  And consequently, the complex problem
presented in this paper reduces to the simple question of: for each
participant, how much more than VCG should we charge?  But, this is
far from straightforward, because (1) it is not clear by what measure
we should define ``as close to incentive compatible as possible,'' and
(2), having chosen a measure, it is typically an intractable
optimization problem to find the optimal prices.

In this paper, we will propose a new way to define \emph{approximate
  incentive-compatibility}, carefully chosen so as to capture the
essential aspects of the design problem, while simultaneously being
computationally tractable enough that the design problem can be
solved.
%

\subsection{Our Solution}
%
%

The goal of this paper, then, is finding good approximately
incentive-compatible prices, with a focus on the CE setting.  In
service of this goal, we begin with a new definition of approximate
incentive compatibility.
Such a definition requires a model for the information set employed by
participants when they devise their strategy
\citep{lubin12approximate}.  Traditionally this is either (1)
\emph{ex-ante}, where the agent knows only the distribution over all
agent values; (2) \emph{ex-interim}, where each agent additionally
knows his own values exactly; or (3) \emph{ex-post}, where all bidders
have complete information.
Instead, we model bidders as being boundedly informed about the values
in the setting.  Specifically, we introduce the concept of a
\emph{blinding distribution}, used to approximate \textit{ex-interim}
prices, while retaining the tractability of \emph{ex-ante} and
\emph{ex-post} formulations.

A key insight of the paper is that when mechanisms are used in
practice, payoffs are generally drawn from a distribution about which
participants have imperfect knowledge.  Thus by moving from the
traditional framework where participants reason about values, to one
where they reason about payoffs (which is, what they ultimately care
about), we obtain a very useful dimensionality reduction which we can
leverage computationally to better be able to minimize the incentive
to manipulate bids.  In this sense, the present paper generalizes VCG
to the more restrictive settings described above.

Based on this new information definition, we then frame the resulting
approximate IC pricing problem as a variational calculus problem, the
solution of which is a novel payment rule for domains where VCG isn't
available.  The full formulation can be interpreted as a Bayesian
Stackelberg game \citep{simaan1973stackelberg}, where the mechanism
designer moves first by picking the parameters of the game to be
played so as to make it as incentive compatible as possible, and the
participants then attempt to obtain as much individual gain as they
can within this game.  The chosen rules and resulting optimal
agent-strategies will be the equilibria of this Stackelberg meta-game.
However, because the game is a stochastic game with an infinite
strategy space it is not tractable to solve for the Stackelberg
equilibrium, and we therefore adopt a simpler Bayes-Nash equilibrium
concept for our solution.

Having constructed this model, we then offer a description of the
space of optimal rules and game equilbria consistent with the
formulation, using variational methods.  This analysis permits us to
critique several existing rules from the literature on a theoretical
basis.  We close by providing multiple examples of numerical solutions
to the variational pricing problem, and characterize these solutions.

In most cases, the rules obtained cannot be described in closed form,
and must be determined computationally.  However, provided suitable
distributions are available to instantiate the information model, the
methods provided can readily be employed to furnish an alternative to
VCG that is as incentive-compatible as possible, and thus as efficient
as possible, given the need to satisfy other hard constraints that
preclude the direct use of VCG.

\section{Theory}



Many payment rules for combinatorial mechanisms have been proposed in
the literature.  We briefly review these, by way of motivating our
method for achieving approximate incentive compatibility.

\citet{Parkes2001threshold} provide the \textbf{Threshold} rule, which
minimizes the maximum \emph{ex-post} regret across agents.  Several
basic rules are also defined therein: \textbf{Small}, which gives all
surplus to those with little of it; and \textbf{Large}, which does the
opposite, among other rules.  Each of these rules is optimal relative
to a specified metric of payoffs at reported values.  However, the
analysis is not in BNE, and thus it is not clear which rule actually
yields the best incentives or the highest efficiency when agents
strategize.
\citet{lubin2009quantifying} propose a \textbf{Reference} rule, that
seeks to minimize the KL-Divergence between the distribution on
payoffs in the reference (e.g. VCG) and in the implemented mechanism,
and show that rules with payoff distributions similar to the reference
will have similar incentive properties in equilibrium.  The work also
evaluates the Threshold, Small and Large rules in approximate BNE, and
finds that among the rules tested, Small often works well.  However,
Small gives no discount at all to those bidders who get the most under
VCG, and is thus hard to argue for in practice.  Further, the 2009
paper evaluates only a small set of fixed rules.  The present work
shows, on the other hand, that we can do better by considering a
broader set of possibilities, especially rules with no closed form
that must be determined computationally.

\citet{erdil2010marginal} define a rule that minimizes the marginal
incentive to deviate from truthful bidding over small misreports,
arguing that this is a minimal condition for approximate
strategyproofness.  However, paper is silent on what to do when larger
misreports are optimal for participants, given the nature of the
underlying domain and design constraints.  By contrast, the present
work handles situations where even very large deviations from truthful
bidding are strategically elicited in the best responses of bidders.

A different thread of work has sought not closed-form rules, but rules
that can be calculated algorithmically, as proposed here.
\citeauthor{conitzer2004amd}'s \emph{Automated Mechanism Design}
\citeyearpar{conitzer2004amd} captures the full design, including
allocations and payments, within this numerical framework, but can
only solve small problems, given the high computational complexity.
\emph{Empirical Mechanism Design}
\citep{vorobeychik2006empirical,vorobeychik2007constrained}
parametrizes the design space and then uses computational techniques
to find optimal parameter settings.  This is similar in spirit to the
approach taken herein.  However, rather than fitting the parameters of
specific functions, we instead formulate our version of the payment
problem in the \emph{calculus of variations} and, lacking a
closed-form solution, perform computation from there.  (Variational
calculus goes all the way back to Bernoulli, but the major development
was by Euler and Lagrange in the 18th century; a good overview of
variational methods is provided by Smith
\citep{smith1998variational}.)

Finally, we point the reader to recent work by
\citet{nisan2011bestresponse}, which identifies auction settings
where, under certain informational assumptions, player best-response
dynamics will lead to the efficient outcome even without fully IC
prices.  One view of the present work is as a generalization of this
idea to settings that are only approximately IC.  Another approach to
avoiding complexity is to create a \emph{replication economy} by
increasing the number of players and goods in order to ``wash out''
local effects.  This is the method employed by the related method of
\emph{Strategyproofness in the large}
\citep{azevedo2012strategyproofness}; by contrast, we reason about the
unmodified economy directly.

\subsection{Preliminaries}
\newcommand{\sni}{\text{-}i}

Although our ideas are more broadly applicable, for simplicity we will
here focus on Combinatorial Exchanges (CEs).  Formally, we have a set
of $\{1..m\} \in M$ goods and $\{1...n\} \in N$ agents.  Each agent
has an initial endowment $E_i \subseteq M : \bigcap E_i = \emptyset$,
and a true value $v_i(\lambda_i)\in \mathbb{R}$ for each potential
trade $\lambda_i \in \mathcal{P}(M)$ of goods they might buy or
sell.\footnote{Without loss we will assume unique goods; all points
  generalize to the case of multiple identical items.}  We will denote
those goods bought $\lambda_i^B$ and those sold $\lambda_i^S :
\lambda_i^B \cup \lambda_i^S = \lambda_i$, and the vector of such
trades as $\mathbf{\lambda} \in \mathbf{\Lambda}$.  Because of the
\emph{Revelation Principal} \citep{myerson1979incentive}, we may
restrict ourselves to mechanisms where agents state a claim of value
(bid) on a potential bundle; however, we must capture that they may
still choose to report untruthfully as $\hat{v}_i(\lambda_i) \neq
v_i(\lambda_i)$.  To simplify notation, we will refer to $V(\cdot) =
\sum_i v_i(\cdot)$ and $\hat{V}(\cdot) = \sum_i \hat{v}_i(\cdot)$; we
use $V_{\sni}(\cdot)$ and $\hat{V}_{\sni}(\cdot)$ for the value of all
agents but $i$.
We are concerned with mechanisms that implement the efficient outcome
at reports, e.g. $\argmax_{\mathbf{\lambda} \in \mathbf{\Lambda}}
\sum_{i \in N} \hat{v}_i(\lambda_i)$ subject to feasibility 
$\sum_{i \in N} \I(g \in \lambda_i^B) \leq \sum_{i \in N} \I(g \in
\lambda_i^S) \; \forall g \in M$, and restricted endowment
$\lambda_i^S \subseteq E_i$.
We denote such an efficient trade as $\mathbf{\lambda}^\ast$, and with
a slight abuse of notation, $v^\ast_i(\hat{v}_i)$ as the value to the
agent $i$ of the efficient trade when he reports $\hat{v_i}$, and
similarly for $V^\ast(\hat{\mathbf{v}})$.

Our mechanism will also charge each agent payments $p_i$, and we note
that these may be negative (for e.g. sellers or swappers) in a CE
setting.  Without loss, we can describe prices by a \emph{discount}
instead, where $p_i = \hat{v}_i(\lambda_i^\ast) - \Delta_i$ for some
$\Delta_i$ specified by the \emph{payment rule}.
We will consider settings of quasi-linear utility; e.g., an agent's
utility or profit is given by $\pi_i(\lambda) = v_i(\lambda_i) - p_i$.
We will sometimes parametrize this as $\pi_i(v_i, \hat{v}_i,
\hat{\mathbf{v}}_{\sni})$, or the agent's profit when he has true value
$v_i$, reports $\hat{v}_i$, and all other agents' reports
$\hat{\mathbf{v}}_{\sni}$.

The well known VCG mechanism uses as its discount the marginal impact
of the agent, e.g. the difference in social welfare between the main
economy and the economy where the agent is omitted.  Formally:
$\Delta_i^{VCG} = \hat{V}^\ast(\hat{\mathbf{v}}) -
\hat{V}_{\sni}^\ast(\hat{\mathbf{v}}_{\sni})$.
We note that the VCG discount enables us to find the \emph{critical
  value}, $v_i^C(\lambda_i^\ast) = v_i(\lambda_i^\ast) -
\Delta_i^{VCG}$, which is the minimum agent $i$ could have bid while
still winning $\lambda_i$ when all $\mathbf{\hat{v}}_{\sni}$ are held
fixed.
This choice of payment is \emph{individually rational}
($\pi_i(\lambda_i^\ast) \geq 0 \forall i$), and \emph{strategyproof}
(agents have a dominant strategy to report $\hat{v}_i=v_i$).  However,
as mentioned in the introduction, in a CE setting VCG is not (weak)
budget-balanced in Bayes-Nash Equilibrium (BNE) ($\exists
\hat{\mathbf{v}} : \sum_i p_i \leq 0$, requiring the center to
subsidize the outcome).\footnote{ A consequence of the
  Myerson-Satterthwaite Impossibility theorem
  \citep{myerson1983efficient}.}

\subsection{Defining Approximate Incentive Compatibility}
\label{sec:defApproxSP}

As was mentioned in the introduction, we want a mechanism consistent
with this CE setup that implements the efficient allocation at
reports, and which uses a payment rule that is individually rational
and budget-balanced.  But we know that to do this we will have to give
up the property of incentive compatibility.  Consequently, we seek a
rule that yields maximal efficency at the reported values in practice.
To do this, we will want the ratio of total value in the allocation at
reported values and at true values to be as close to one as possible,
e.g. $\nicefrac{\mathbf{V}(\hat{\mathbf{\lambda}})}
{\mathbf{V}(\mathbf{\lambda}^\ast} \rightarrow 1$.  A rule that yields
the closest such limit, we deem to be approximately incentive
compatible, and is our overall goal.
More concretely, we need to specify: (a) what types of misreporting a
participant can perform and (b) what information the participant has
when deciding how to misreport (i.e. the nature of the BNE).  For the
setting outlined in the previous section, (a) is simple: we allow for
all possible misreports $\hat{v}_i(\lambda_i)$; e.g., particupants may
state some other value than the truth for all bundles $\lambda_i$.
The information-set (b), though, requires explication.

\subsubsection{Participant Information-Sets}
\label{sec:infoSets}

Because we are doing our analysis in full BNE, if the information-set
is complex, then evaluating equilibrium is likewise computationally
complex, and often prohibitively so.  Consequently a number of
proposals have been made in the literature for simple information-sets
that lead to tractable analysis of the degree of misreporting, making
the problem of finding the degree of efficiency loss solvable, and
thus making design under the given approximate IC criterion
possible. (See Appendix~\ref{app:existingApproxIC} for details, and
\citet{lubin12approximate} for a longer discussion.)

The simplest approach is to assume that participants are \emph{fully
  informed} about all values in the market.  Because this sort of
analysis is using information that is typically available only after
the mechanism is complete (even for decisions that are made up front),
it is often referred to as an \emph{ex-post} analysis.  Consequently,
this choice of information-set collapses our BNE calculation to a
normal Nash equilibrium.  While appealing for simplicity and
computational reasons, it is highly unrealistic in many settings.

Instead, one can assume that participants make their strategic
decisions when they have no more than an expectation as to the values
in the market.  In such an \emph{ex-ante} model, the participant has
only probabilistic information about \textit{his own} value.  This is
unrealistic in most settings -- e.g., while \emph{ex-post} is too
informed, \emph{ex-ante} is too \emph{un}informed.
Consequently an \emph{ex-interim} information-set -- where the
participants know their own value, but have only probabilistic
information about all the values of other bidders -- attempts a
``Goldilocks'' formulation between the other two.  However, such a set
is computationally challenging compared to the other two, because it
possesses neither the simplification of full information, nor of
everything being performed in a simple expectation.

\subsubsection{Simplification Via the Potential Profit}
\label{sec:simpCrit}


One contribution of this work is a novel information-set definition
that has the tractability of an \emph{ex-post} measure, but the
economic intuition of an \emph{ex-interim} measure.

In considering what information-set to use, we focus on the
information a participant both needs and has in making his decision
about how to interact with the mechanism.  In practice, bidders are
boundedly informed and boundedly rational, and we want a model that
captures this.

It seems reasonable to assume that participants know not only their
own value for a bundle of goods, but also what they will pay for a
bundle, conditioned on bidding for it and winning it.  Further, that
they know they may lose the bundle, and in fact may have an idea of
how likely this is, as a function of their bid report.

Dispite its common use in analysis, it is generally unreasonable to
assume participants know the exact (\emph{ex-post}) values of other
bidders for all bundles, or even for the allocated bundle.  However,
firms often have some idea of their opponents' business plans -- after
all, they are typically in the same industry.  And thus they may have
distributional information about the values of their opponents.  That
is, they may be able to view their competition as having been sampled
from some IID population of competitors.

However, as it turns out, we can conveniently summarize all of this
information into a single unidimensional property: the distribution on
available profit under the VCG rule to a given participant.  As we
shall see, it is this distribution that really captures the
competitive environment a participant finds himself in.
Formally we have the following (proofs are left to
Appendix~\ref{app:obs}):


\begin{observation}
\label{ob:PSI}
  Holding the bids of the other players $\hat{\mathbf{v}}_{\sni}$
  constant, each winning bid $\hat{v}_i$ has two additive
  components: (1) an amount $v_i^C$ that by definition is needed or
  the bidder will lose his bundle and (2) an amount $\psi \geq 0$ in
  excess of this.
\end{observation}
Observation~\ref{ob:PSI} lets us focus, not on the joint trade and
reported valuation profile when considering a bidder's bid (and
subsequent payment), but instead solely on the amount $\psi$, the
amount over the critical value that is offered in the bid.  $\psi$ is
strictly positive, because were it to ever run negative, the bidder
would have reported below its critical value, and lost the bundle.

\begin{observation}
\label{ob:VCGPSI}
  For VCG, $\psi = \pi_i^{VCG} = \Delta_i^{VCG}$ in equilibrium (e.g.
  the VCG profit is the discount in equilibrium). 
\end{observation}
Observation~\ref{ob:VCGPSI} lets us give $\psi$ a useful name and
interpretation: the bidder's profit as calculated by the VCG Rule (but
in general, not necessarily at truthful bids of others).  In VCG,
where $i$ is truthful, this will be equivalent to the VCG discount.
Alternatively, we can view $\psi$ as the amount of surplus
attributable to $i$ at his reported value, be that truthful or
otherwise.  We refer to $\psi$ as the \emph{potential profit},
i.e. the amount VCG would offer, even if our eventual rule often will
not be able to provide this much and maintain its other requirements
(such as budget balance).

In general, $\psi$ should be consided to be taken relative to the
\emph{reported} values of other players, i.e. when they behave in
equilibrium.  But it is calcuated at the true value of the player
whose incentives are being considered.  That is, as $\psi =
v_(\lambda_i) - v_i^C$, the first term is $i$'s truthful value, while
the second term is $i$'s critical value, which is independent of $i$
and is taken at the equilibrium reports of the other agents.  In other
words, it is the amount an agent would make if it told the truth,
everyone else played the equilibrium, and the mechanism was charging
this particular bidder as if it were VCG. As we shall see, this
unusual structure enables us to capture both $i$'s true value and the
environmental conditions $i$ faces simultaneously.

\begin{observation}
\label{ob:outEq}
  Out of equilibrium but above $v_i^C$ the reported discount $\psi =
  \hat{\Delta}_i^{VCG}$ (e.g. the VCG formula evaluated at a
  non-truthful bid) shifts by exactly the amount of the misreport,
  such that the true profit remains the same, $\hat{\pi}_i^{VCG} =
  \pi_i^{VCG}$.
\end{observation}

Every closed-form payment rule reported in the literature to date,
including VCG, Threshold, Small, Large, etc, can be expressed as
$\Delta(\psi)$, the discount supplied when the report is $\psi$ above
$v_i^C$ \citep{Parkes2001threshold}.  Thus $\psi$ is a convenient way
to parametrize a payment rule, and for the same reasons it is a
convenient way to think about the information available to a bidder.
For example, VCG has a characteristic signature when viewed from this
perspective: the discount supplied rises one-for-one with $\psi$,
which Observation~\ref{ob:outEq} reveals to have the effect of keeping
the profit constant regardless of the report.  It is this that
ultimately makes VCG strategyproof.  Non-incentive compatible rules
won't have this property.  But by focusing on profit under a given
rule as a function of $\psi$, we capture bidders' incentive structure
perfectly, conditioned on the behavior of other players and upon the
allocation remaining fixed.


\subsubsection{The Distribution of Potential Profit}
\label{sec:distribution}




Because we want to do design in BNE, we must be able to reason about
the distribution on possible participant values, rather than a
specific outcome.  Consequently, we let $f(\psi)$ represent a
distribution over the potential profit that the bidder would obtain if
he were paying the VCG rule, given the reported behavior (under the
actual mechanism being used, not VCG) of the other bidders, and
conditioned on winning $\lambda_i^\ast$. It is thus dependent on the
reports of the other agents, as well as dependent on the true value of
agent $i$.

The distribution is over many counterfactual instantiations of the
market domain drawn IID from some consistent underlying generation
process.
In the real world, such a distribution may be built up by considering
the many examples in a repeated game if the dynamic process is
reasonably stationary.  Alternatively, one can model the participants
and view the distribution as having been constructed from many
instances drawn from a synthetic bid generator.

Our model is consistent with the distribution over the rest of the
agent's value either occurring at truth, or having been calculated
according to the BNE of the resulting overall market mechanism.  In a
synthetic context, using the truth-based distribution will be easier,
as finding the reported values of the other agents would require
solving for the BNE of the mechanism.  However, in a repeated
real-world game, the observed bids are presumably approximately
equilibrium bids (depending on the sophistication of the players),
making the equilibrium-based distribution the easier version to
obtain.  Our rule is not agnostic to this distinction: it will provide
different rules depending on the calculated $f$.  It is, however,
well-defined and consistent for both.  We will focus on the version
where the other agents are bidding in equilibrium, because we view
this as the more natural instantiation of the mechanism.

\subsubsection{The Blinded Regret Information-Set}
\label{sec:blindedRegret}


If bidders know $\psi$ exactly and they know the payment rule in
place, they have \emph{ex-post} information -- they know exactly the
topology of the incentive structure they face.  This is exactly
equivalent to their knowing their critical value exactly, which is
functionally equivalent to knowing the bid values of all the other
agents in the winning allocation.  They can then behave optimally,
bidding exactly their critical value, which, under any individually
rational payment scheme, should give them maximum profit.  Clearly
this \emph{ex-post} structure is undesirable.  The construction of $f$
alone does not ameliorate this: $f$ tells us how frequently a given
value of $\psi$ occurs, but if agents know this exact value when they
choose their report, we are still in an \emph{ex-post} scenario.  
By contrast, if bidders must choose their report based only on knowing
$f$ in aggregate, but not which particular value of $\psi$ they
directly face, we have an \emph{ex-ante} condition -- the bidder has
no concrete knowledge of his own value.  For the reasons described in
section~\ref{sec:infoSets}, neither of these extremes is desirable.

We would prefer an information-set that behaves more like an
\emph{ex-interim} condition.  Our solution is to adopt an explicit
model of the uncertainty that the bidder has about the value of joint
bids, which we call the \emph{blinding distribution}.
More specifically, given that $f$ represents the actual distribution
of VCG-rule profits available in the mechanism, we use a structure
where bidders only have a guess about what their particular profit,
$\psi$, is, conditional on the overall actual distribution being $f$.  The
blinding distribution $\mu_x(\psi)$ represents the bidder's belief
about the value of $\psi$, conditioned on $x$ being the true
$\psi$. We then compound $\mu_x$ with $f$ to produce
\begin{equation}
\label{eq:defg}
g(\psi) = \int f(x) \mu_x(\psi)dx
\end{equation}
and it is then this function, $g$, which we assume the bidders have
access to, not $f$.  In the experiments in
section~\ref{sec:experiments}, we use a truncated Normal for $\mu_x$.

This approach allows us to smoothly interpolate between an
\emph{ex-ante} condition, which occurs when $\mu_x$ is the uniform
distribution and the expected \emph{ex-post} regret, which occurs when
$\mu_x$ is the Dirac $\delta$ distribution.  Cases between these,
e.g. when $\mu_x$ is Normal, represent an interim information state.
This state is not identical to that defined in the richer
\emph{ex-interim} deviation incentive (see
Appendix~\ref{app:existingApproxIC}, although intuitively they both
lie between \emph{ex-ante} and expected \emph{ex-post}).

\subsubsection{Blinded Regret Approximate Incentive Compatibility}

Given the attractive information-set just described, knowlege of $g$
and of the payment rule in place, we can quantify how much incentive
an agent has to modify his bid away from truth, the
\emph{Blinded Regret Deviation Incentive}:

\begin{equation}
\label{eq:blindedRegretDI}
\text{BlindedRegretDI}_i = \int  
                           \max_{\hat{v}_i}  \left[
                              \pi_i(\hat{v}_i, \tilde{\psi}) \right]
                              g(\tilde{\psi}) d\tilde{\psi}
                           - \int \pi_i(v_i, \psi)f(\psi)d\psi
\end{equation}
where $\tilde{\psi}$ is bidder $i$'s belief about his potential
profit, when he has been blinded by some distribution $\mu$ that has
been compounded with $f$ to form $g$ by
equation~\ref{eq:defg}. Further, $\pi_i(\hat{v}_i, \tilde{\psi})$ is
the true profit to bidder $i$ under the actual rule being evalated,
when he reports $\hat{v}_i$ and his available potential profit is
$\tilde{\psi}$.  The first term then represents the amount the bidder
believes he can profit by best misreporting under the blinded regret
information-set.  The second term is the amount bidder $i$ can profit
if he simply reports the truth in the mechanism.  Thus the difference
is the potential gain of the bidder when he best responds.
Given this definition of approximate incentive compatibility, we next
turn to the construction of optimal rules by this criterion.

\subsection{Optimal Approximately Incentive Compatible Payment Rules}

If we want to computationally obtain a perfectly IC payment rule, we
can cast the problem as an Automated Mechanism Design
\citep{conitzer2004amd}, as we do in Appendix~\ref{app:AMD}.  However,
the full problem is hopelessly intractable for any but the simplest of
instances.  But, by focusing on $\psi$ instead of on the full trade
and allocation space, we have reduced the complexity of the payment
function definition massively.  Contingent on the other bidders being
fixed and on winning the bid, this reduction has been without loss and
is perfectly consistent with all prior closed-form rules in the
literature, and is consequently the approach we take here.
%
%

We will find it convenient to work with functions that move in the
same direction as payments, unlike the discounts used so far.
Consequently, we define $r(\psi) = \psi - \Delta(\psi)$.  This
function represents the amount the agent is asked to pay above the
critical value (which he must alway pay in any reasonable mechanism).
Consequently, the final payment the agent is asked to make will be
$p_i=v_i^C+r(\psi)=v_i^C+r(\hat{v}_i-v_i^C)$.  Further, this function
has a useful interpretation as the ex-post regret a bidder has for
reporting $v_i^C + \psi$ for $\lambda_i$ instead of reporting exactly
the critical value $v_i^C$, which for all IR payment rules is the
optimal report (i.e. $r$ represents the potential gain if the agent
were able to making the \emph{ex-post} optimal report).
We note that all existing closed-form rules treat all bidders
symmetrically, and using $r(\psi)$ enables us to do the same by
applying the same payment rule to all bidders.  Our task, then, is to
identify payment rules $r$ that are optimal under the blinded regret
approximate IC criterion we have just identified.

\subsubsection{The \emph{Ex-Ante} Variational Problem}
\label{sec:constStratTheory}

We first copnsider the case where the blinding distribution $\mu_x$ is
Uniform over the full domain and thus, after blinding, the agent has
no concrete information about his true value.  This case corresponds
to an \emph{ex-ante} condition.  Moreover, we restict ourselves to the
case where the bidder adopts a strategy where he chooses but a single
``shade'' parameter $s \in \mathbb{R}$ by which to offset his true bid
in the mechanism (and we generalize this in the next section).
In this case, the bidder's problem becomes an unconstrained non-convex
optimization in one dimension:
\begin{equation}
  \argmin_s \int
            \left(\I_{\psi\geq s} r(\psi - s) + \I_{\psi < s} \psi \right)
            f(\psi)d\psi
  \label{eq:constAgent}
\end{equation}
where $f(\psi)$ is a distribution over the profit that the bidder
would obtain if he were paying under the VCG rule (and everyone else
is in equilbrium), and $s$ is the amount that the agent shades his
report down from, below his true value, and $\I$ is the indicator
function.

In \eqref{eq:constAgent}, $f$ is used to construct an expectation for
the regret the bidder retains for reporting $v_i^C+\psi-s$ instead of
the ideal $v_i^C$ when $s$ is small enough that he still wins, plus
his regret over the full potential profit in the case where the bidder
has shaded so much that he has lost the bid.  By minimizing this
aggregate regret, the bidder is simultaneously maximizing his expected
reward.%
\footnote{Because we have inverted this optimization for clarity of 
exposition, our problem now appears to be a min-min problem; but note
that due to the structure of the coupling between this and the
subsequent problem, we do indeed still have a min-max problem.}

The structure of the center's problem is specific to the mechanism
being designed, and so we now specialize to budget-balanced CEs.  We
have a constrained linear-variational program:%
\begin{align}
  \argmin_{r} & \int r(\psi)f(\psi)d\psi & \text{Regret at truth}
                                                 \label{eq:constCenter} \\
  \text{s.t.} \,\,\quad\ & \int r(\psi-s)f(\psi)d\psi \geq k & \text{BB}
                                                 \notag \\
                         & 0 \leq r(\psi) \leq \psi \forall \psi & 
                                       \Delta^{VCG} \text{\& IR} \notag 
\end{align}
The variable in this variational program
$r:\mathbb{R}\rightarrow\mathbb{R}$ is picked to minimize the expected
regret the bidder will face when bidding truthfully.
The first constraint ensures that the total regret when the bidder
shades must be at least $k$.  For a suitably chosen $k$ this ensures
that a sufficiently small amount of discount be doled out, such that
budget balance can be achieved.
Next we constrain the rule to ensure that no discount is negative (or
equivalently, $r(\psi)\leq\psi$), as a negative discount implies you
must pay more than your report, violating individual rationality for
truthful bids.
Lastly, we include any additional constraints from the setting we
need, as additional restrictions on $r$.  Here we include the
restriction that discounts must be no greater than VCG (or
equivalently, $r(\psi)\geq 0$), to be consistent with the other rules
from the literature.
We will return to this setup in Section~\ref{sec:experiments} to
illustrate its solution in a typical environment.

\subsubsection{The Blinded Regret Variational Problem}
\label{sec:funcStratTheory}

Now we turn to realizing the more complex case where we both have a
blinding distribution, and where the bidder's strategy is defined as a
function $s(\psi)$, rather than a simple constant.
With this change we obtain:
\begin{equation}
  \argmin_s \int
            \left(\I_{\psi\geq s(\psi)} r(\psi - s(\psi)) + 
                  \I_{\psi < s(\psi)} \psi \right)
            g(\psi)d\psi
  \label{eq:funcAgent}
\end{equation}
Note that in this formulation as we let $\mu_x \rightarrow \delta$ we
will move into an expected \emph{ex-post} regret world and the bidder
will be able to respond optimally for every $\psi$.  The solution then
becomes $s(\psi)=\psi$, with the bidder reporting exactly $v_i^C$
everytime.  We thus ensure that we always choose $\mu_x$ such that it
always has a non-zero variance.

In concert with this change for the bidder, a few updates are in order
for the center as well.
First, if the bidder is reasoning based on partial information, it
makes sense for the center to use $g$ in its objective also, so it is
attempting to incentivize the agent based upon the information-set
that the bidder will ultimately use.
Second, the assumption that the center has a perfect model of the
bidder's strategy is also too strong.  To relax it, we adopt a
blinding distribution $w_x(\psi)$ for the center as well, and apply
the resulting compound distribution $h(\psi)$ in the budget-balance
constraint.  $h$ can either model uncertainty on the center's part
about $f$, or equivalently, uncertainty about the choice of $s(\cdot)$
that the bidder will employ.  Both are available from the same
formulation because $s$ can be moved into the argument of the
distribution function in place of $r$ by a suitable change of
variables in the integration, making the two interpetations valid upon
the same expression simultaneously.  Putting this together, we obtain:
\begin{align}
  \argmin_{r} & \int r(\psi)g(\psi)d\psi & \text{Regret at truth}
                                                 \label{eq:funcCenter} \\
  \text{s.t.} \,\,\quad\ & \int r(\psi-s(\psi))h(\psi)d\psi \geq k & \text{BB}
                                                 \notag \\
                         & 0 \leq r(\psi) \leq \psi \forall \psi & 
                                  \Delta^{VCG} \text{\& IR} \notag 
\end{align}
The use of blinding distributions significantly improves the realism
of our model, moving us solidly into a partial information setting.

%

\subsubsection{Analysis via Variational Methods}




Our formulation has provided us with several variational optimization
problems.  Our first step is to construct the necessary boundary
conditions.  Our restriction that $0 \leq r(\psi) \leq \psi$, forcing
the rule to be IR and within the discount envelope of VCG, provides a
necessary such condition.

With these in place, we might ideally be able to use the
Euler-Lagrange equation to transform our optimization problems into
sets of differential equations and then solve these for the optimal
functions \citep{wan1995variational}.  But because, Euler-Lagrange can
only be applied where we have a non-linear variational form involving
at least one set of derivatives, a general closed-form solution is not
available by this path.

However, the literature on Variational forms \emph{does} extend to
this case, and offers us the following important insight
\citep{wan1995variational}: We should expect the function that is the
solution to problems with our structure, to always lie on one bound or
the other (e.g. the optimal $r(\psi)$ will either be $0$ or $\psi
\forall \psi$ pointwise).  This immediately shows that rules such as
the current state-of-the-art Threshold rule, that take on intermediate
values, will not be optimal under our criteria.  Rules such as Small
or Large are admitted in the class, but the class is far larger than
just these existing rules.
Additionally, if we know the shape that $r$ should take (e.g. Large),
we can in fact use variational methods to solve for the parameters of
this shape.
%

\subsubsection{Characterization as a Stackelberg Meta-Game}


We note that the formulations provided in the previous two sections are
are in fact min-max functional programs: the center first minimizes
over payment functions the expected deviation incentive functional;
but this functional is itself a maximization over potential bidder
misreports of the bidder's expected gain in profit over being
truthful.
Because the two problems are inter-dependent and push in opposite
directions, they can be interpreted as a 2-player game.  Specifically,
since there is a distinct first-mover in the form of the center, who
must credibly declare the rules it will use before the bidders then
participate, formally we have a Bayesian Stackelberg game. See
\citet{basar1995dynamic} for details on Stackelberg games.
This is, in fact, a general property of the task of \emph{mechanism
  design}.  We can view the design process itself as a Stackelberg
game: the designer/center chooses a mechanism that provides properties
she likes best.  The players who then participate within the chosen
design move second and attempt to maximize their own reward, within
the structure they have been given.  In otherwords, there is a
Stackelberg meta-game whose outcome determines not only play in the
eventual actual game (the second move), but also the choice of game to
be played itself (the first move).

Stackelberg games are typically solved by backwards induction, where
we solve for the optimal behavior of the second-mover first, and then
find the first-mover's strategy that would maximize his reward given
the behavior of the second player.
Given that moves in our meta-game are the choice of a continuous
one-dimensional function ($r$ and $s$ respectively), such a backward
induction process is prohibitively expensive, even after the above
simplifications we have employed.
So rather than trying to directly solve the min-max stochastic
functional program needed to find the outcome of the Stackelberg
meta-game directly, we recast the problem as two separate and opposed
optimization problems: one for the bidder's goal of maximizing his
expected profit, and another for the center's goal of minimizing
incentives to misreport.
This enables us to solve for the outcome of the BNE of the meta-game
with traditional iterated best-response methods.  The result, though,
will be a BNE of the meta-game, not necessarily a Stackelberg
equilibrium of the meta-game.  The true Stackelberg equilibria will be
those BNE that have maximum value for the first mover (in our case the
center).  In our numerical experiments, we have not been able to
identify more than one stable equilibrium, and so we report on the
single equilbria we have identified.

\section{Methods}


%

\subsection{Parametric Distributions for $f$, $\mu$ and $w$}




While it is entirely possible to define $f$ by an empirical
distribution, in practice it will often be convenient to use a
parametric form instead.
Because $V^\ast$ and $V^\ast_{\sni}$ are constructed from a
maximization process, their distributions are well-modeled by
Generalized Extreme Value (GEV) distributions
\citep{coles2001introduction}.  We are interested in the distribution
of the difference in these values, $\Delta_i^{VCG}$ for a truthful
bid, because by Observation~\ref{ob:outEq}, this is equal to the VCG
payoff.  Because the difference of \emph{uncorrelated} GEV variables
is a Logistic distribution, one might think that a Logistic would be
appropriate.  However, these variables are \emph{highly} correlated.
Intead it is more appropriate to think of $V^\ast$ as an
\emph{exceedence} over the threshold of $V^\ast_{\sni}$.  A model of
this form results in a Generalized Pareto Distribution (GPD)
\citep{pickands1975statistical} for $f$.  This is consistent with
previous work that has fit the distribution of VCG payoffs in a CE
setting empirically, and found that it well matched the GPD
\citep{lubin2009quantifying}.%
\footnote{%
We note that that the exponential distribution is a special case of
the GPD (where shape parameter is 0).  As shown in
Appendix~\ref{app:ratio}, the ratio of $f(\psi-s)$ to $f(\psi)$ drives
the determination of our eventual payment rule.  Because the
exponential distribution has a fixed \emph{failure rate}, the ratio of
fixed offsets of its pdf is a constant.  This means that all feasible
rules are possible solutions when $f$ is exactly an exponential, under
the simpler \emph{ex-ante} model of
Section~\ref{sec:constStratTheory}.  This is not true for the more
complicated subsequent model.%
}%
We also note that the GPD is a limiting case model, and other
distributions are certainly possible in practice.  We therefore also
consider the Burr XII distribution which generalizes the GPD and the
Log-Logistic model \citep{burr1942cumulative}; it is an attractive
choice as it has been used to model income distributions, which is
effectively what the VCG profit represents
\citep{champernowne1952graduation}.

In our experiments, we use simple truncated normal distributions for
both $\mu$ and $w$.  More complex models of bidder information could
easily be incorporated, but in the absence of a specific context, we
opt for a simple distributions.

\subsection{Finding the BNE of the Meta-Game}

Ultimately, we will need to solve for the equilibrium of our game.
For finite domains, min-max programming techniques might suffice
\citep{aissi2009min}.  However, our problem has infinite support, and
thus we employ \emph{iterated best response} techniques, which go all
the way back to \citet{cournot1897researches}.  Specifically, we
proceed in rounds where we find the best response for each player to
the strategy employed by their opponent in the previous round.  For
recent treatments of such techniques in settings similar to our own,
see Reeves et. al. and Vorobeychik et
al. \citeyearpar{reeves2004computing,vorobeychik2008stochastic}.  Our
numerical approach employs a damped form of this method, i.e. with
fictitious play \citep{robinson1951iterative,brown1951iterative}.  As
Shapley noted \citeyear{shapley1964some}, this procedure may fail to
converge with cycles of best-responses repeating themselves.  But, we
have been able to force convergence in most cases by imposing suitable
(and reasonable) additional boundary conditions, and by leveraging the
numerical advantages that happen to come along with the smoothing
associated with the \emph{blinding distributions} described in
section~\ref{sec:blindedRegret}.


Next, we describe how we find the best response for each of our
variational problems by an appropriate numerical method:

\subsection{The Bidder Problem}

For the optimal strategy formulation of
section~\ref{sec:constStratTheory} where $s$ is a constant, we need
only solve a 1-D non-convex optimization to find $s \in \mathbb{R}$.
Because the problem is unconstrained, it suffices to employ a Brent
Line Optimizer \citep{brent2013algorithms}.  We note only that it will
often be the case that there will be multiple minima with near-equal
values.  The numerical stability of the algorithm is improved if such
ties are broken in a consistent way, e.g. by always picking the minima
closest to zero (which is a reasonable assumption for bidder behavior
anyway).  To implement this, it suffices to perform a simple grid
search for potential minima, local search via the Brent solver to
improve these results, and then pick the closest to $s=0$ within the
equivalence class of minimal solutions.
In the more complex formulation of section~\ref{sec:funcStratTheory}
where $s(\psi)$ is a function, we simply discretize $\psi$ into $B$
bins and run the above method on each.  We then approximate $s(\psi)$
as a simple linear interpolation over these points.
%

\subsection{The Center Problem}
\label{sec:methodCenter}

The Center's problem is more complicated as a constrained variational
program.  Nonetheless, if again we discretize $\psi$ into $B$ bins, it
is straightforward to then cast the problem as an LP that can be
solved with e.g. CPlex.  The standard version
(section~\ref{sec:constStratTheory}) of the problem is then as
follows:
\begin{align}
  \min_{r_b : b \in B} & \sum_{b \in B} 
                              (F(\overline{b}) - f(\underline{b})) r_b
                                          &  \text{Regret at truth} \\
  \text{s.t.}\quad\quad& \sum _{b \in B}
                              (F(\overline{b}) - F(\underline{b}))
                              r_{b-s(\text{\sout{$b$}})} \geq k  
                                          &  \text{BB} \notag \\
                       & 0 \leq r_b \leq \text{\sout{$b$}} \forall b \in B
                                          &  \Delta^{VCG} \text{\& IR}  
                                             \notag
\end{align}
Where $\underline{b}$, \sout{$b$} and $\overline{b}$ are the lower,
mid and upper points of bin $b$ respectively, and where suitable
boundary conditions are in place to ensure that the indexing stays
within range regardless of the $s$ function.
The more complex version of the problem is identical but uses the
compound $h$ distributions instead of $f$.
In both cases we ultimately construct a continuous version of $r$ from
the discrete by simple linear interpolation.

As it turns out, there is also a greedy method available that solves
the above problem exactly.  In the interest of space, we defer the
details to Appendix~\ref{app:ratio}.

%
%
%
%

\section{Experiments}
\label{sec:experiments}

To validate this approach, we have conducted a series of experiments,
using parametrized distributions for $f$.  Leveraging the methods
described in the previous section, we can quickly compute the
necessary equilibria.  With suitable configuration, all experiments
converge in less than 50 rounds and run in a few minutes on an Intel
Core i5.  In all of the experiments that follow we have used a
discretization of 50 bins and 200-fold sub-sampling when evaluating
integrals over the bins.  In all cases, we evaluated $\psi$ over a
range $[0,10]$.  Consequently all distributions were truncated such
that their support did not exceed this range.

\subsection{\emph{Ex-Ante}, $f$ $\sim$ Pareto}

\begin{figure}[tb]
\centering
\begin{minipage}[t]{.48\textwidth}
  \centering
  \includegraphics[width=\textwidth]{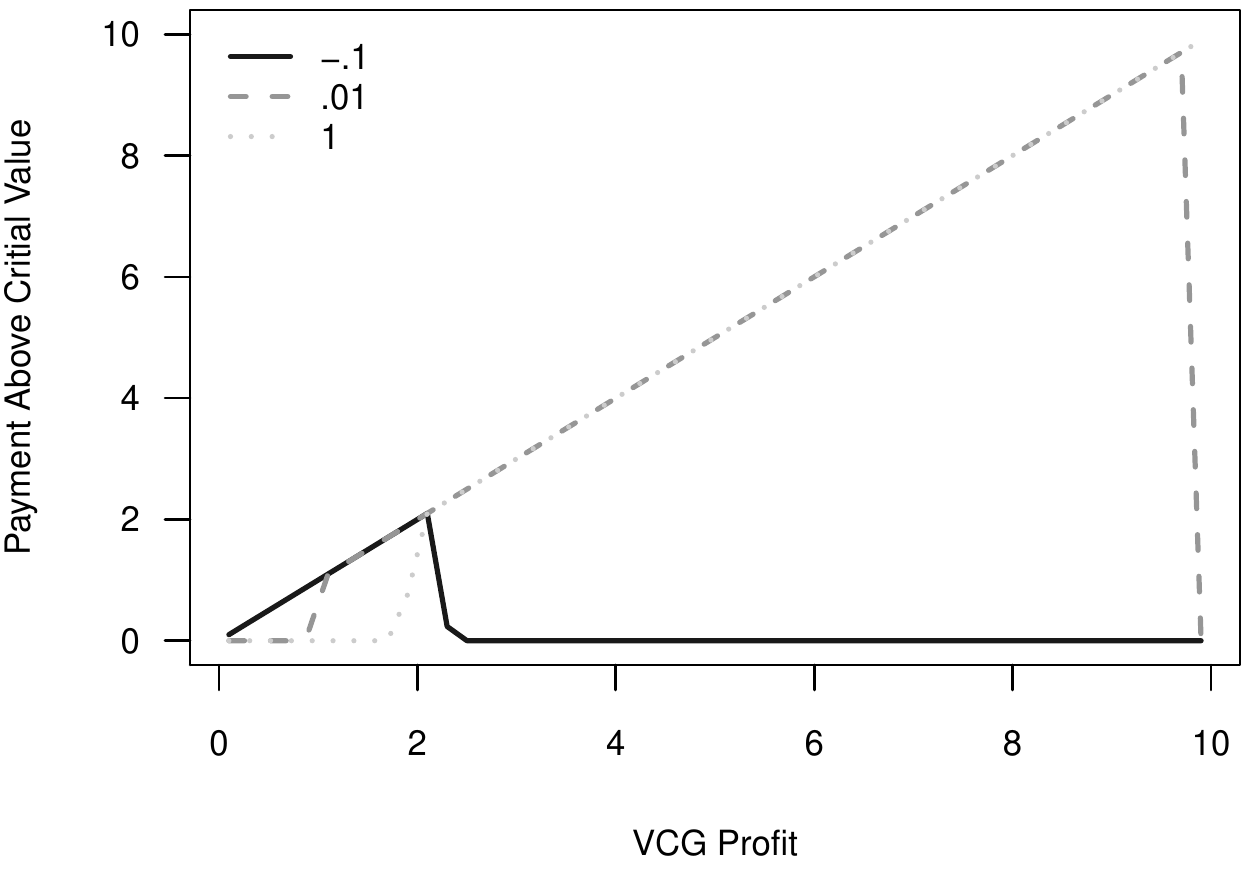}
  \caption{\emph{Ex-ante} payments above the critical value for
    several examples of the Pareto shape parameter.}
  \label{fig:constantParetoShapeRegret}
\end{minipage}%
\hfill
\begin{minipage}[t]{.48\textwidth}
  \centering
  \includegraphics[width=\textwidth]
{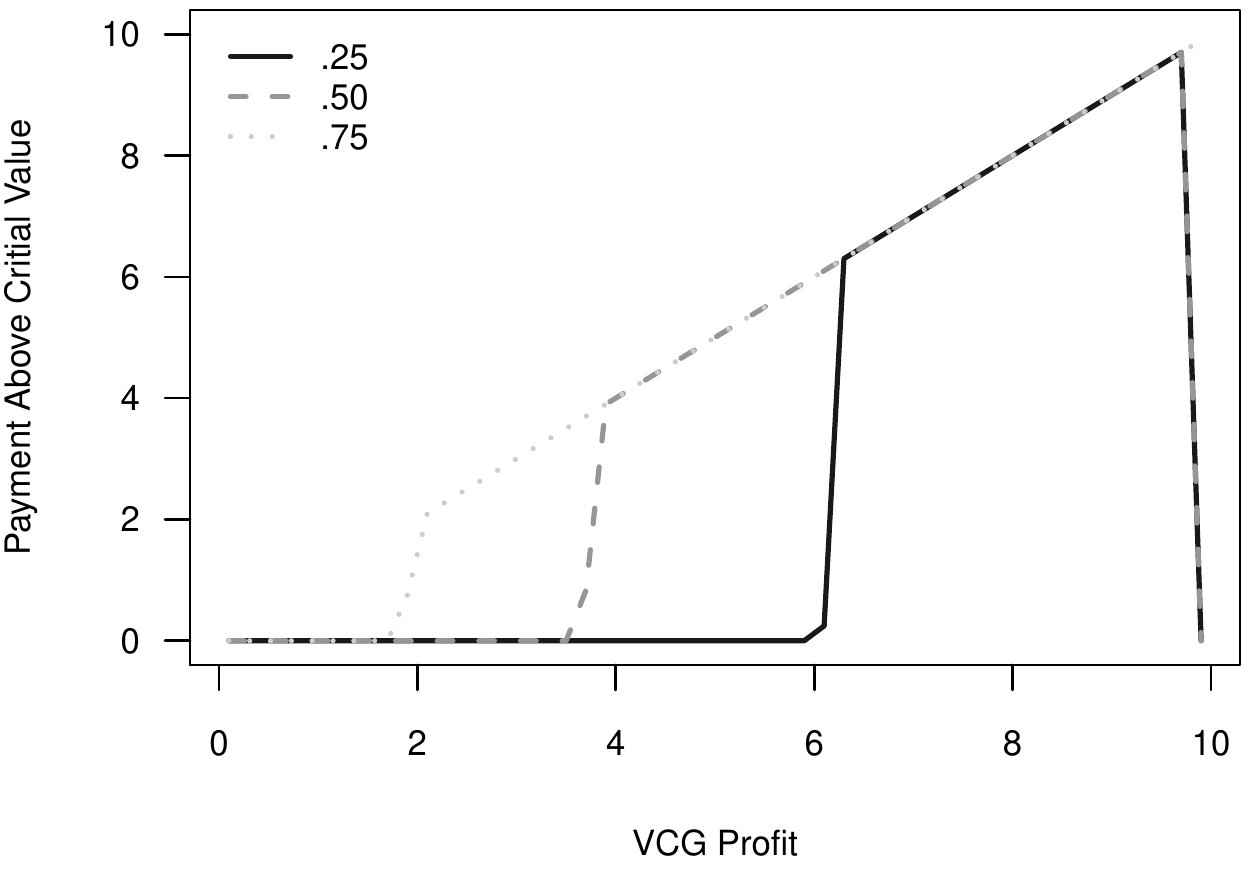}
  \caption{\emph{Ex-ante} payments above the critical value for a 
           Pareto distribution, varying 
           the fraction of required regret, i.e. $\gamma=\{.25,.50,.75\}$}
  \label{fig:constantParetoReqRegretRuleRegret}
\end{minipage}
\end{figure}

For the first experiment we used the basic \emph{ex-ante} formulation
from section~\ref{sec:constStratTheory}.  We equipped it with a
Generalized Pareto Distribution for $f$ with position $0$, and scale
$1$.  We then varied the shape parameter in $\{-.1, 0.01, 1\}$.  We
find that the optimal strategy for the bidder under all three settings
is to shade by $0.2$.
The rules produced under this setting are shown in
Figure~\ref{fig:constantParetoShapeRegret}.  For a shape parameter of
$-.1$, the calculated rule charges only the bidders with small $\psi$.
Consequently, the rule selected in this case is the Large rule, which
gives all the available surplus to bidders with large potential
VCG profit.
By contrast, when the scale parameter is $1$, a heavy-tailed
distribution, the obtained rule asks only those bidders with the
largest $\psi$ to pay.  Accordingly, this parametrization selects the
Small rule, which gives all the available surplus to bidders with the
smallest VCG potential profits.
In previous work \citep{lubin2009quantifying}, it was found that the
GPD well-matched empirical data, and that the Small rule was
observed to perform exceedingly well in restricted BNE.  From these
figures, we can see why.  If the GPD distribution has a positive shape
parameter, then the optimal rule, under our model, is likewise
Small.

\subsection{\emph{Ex-Ante}, $f$ $\sim$ Pareto For Different Budgets}

The budget balance constraint described in
section~\ref{sec:constStratTheory} enforces that the rule cannot dole
out more surplus than is available in the domain, $k$.  VCG violates
this and gives out enough surplus to achieve full strategyproofness,
an amount we can quantify as: $k^{VCG} = \int \psi f(\psi) d\psi$.
Because of this it is convenient to consider reparametrizing in terms
of this quantity and focus on the constant $\gamma = \frac{k}{k^{VCG}}
\in [0,1]$.  $\gamma$ represents the fraction of the total VCG surplus
that must not be given away if we are to achieve budget balance.  All
things equal, the larger $\gamma$, the more the agents are going to
have to pay.
This can clearly be seen in
Figure~\ref{fig:constantParetoReqRegretRuleRegret}, which shows the
bidder payments above the critical value where $\gamma \in \{.25, .50,
.75\}$.  Further, the more surplus there is to hand out, the better
the rule can do in minimizing the incentives to deviate, and thus the
optimal strategy falls from shaving by $0.2$ when $\gamma=.75$ to
$0.1$ when $\gamma=.25$.
Addtional results based on the \emph{ex-ante} formulation, are
provided in Appendix~\ref{app:exanteBurr}.

\subsection{Blinded-Regret and Functional Strategy, $f$ $\sim$ Pareto}
\begin{figure}[tb]
  \centering
  \subfigure[$\mu\sim$ Normal(0,2)]{
\vspace{-1em}
    \includegraphics[width=.45\textwidth]{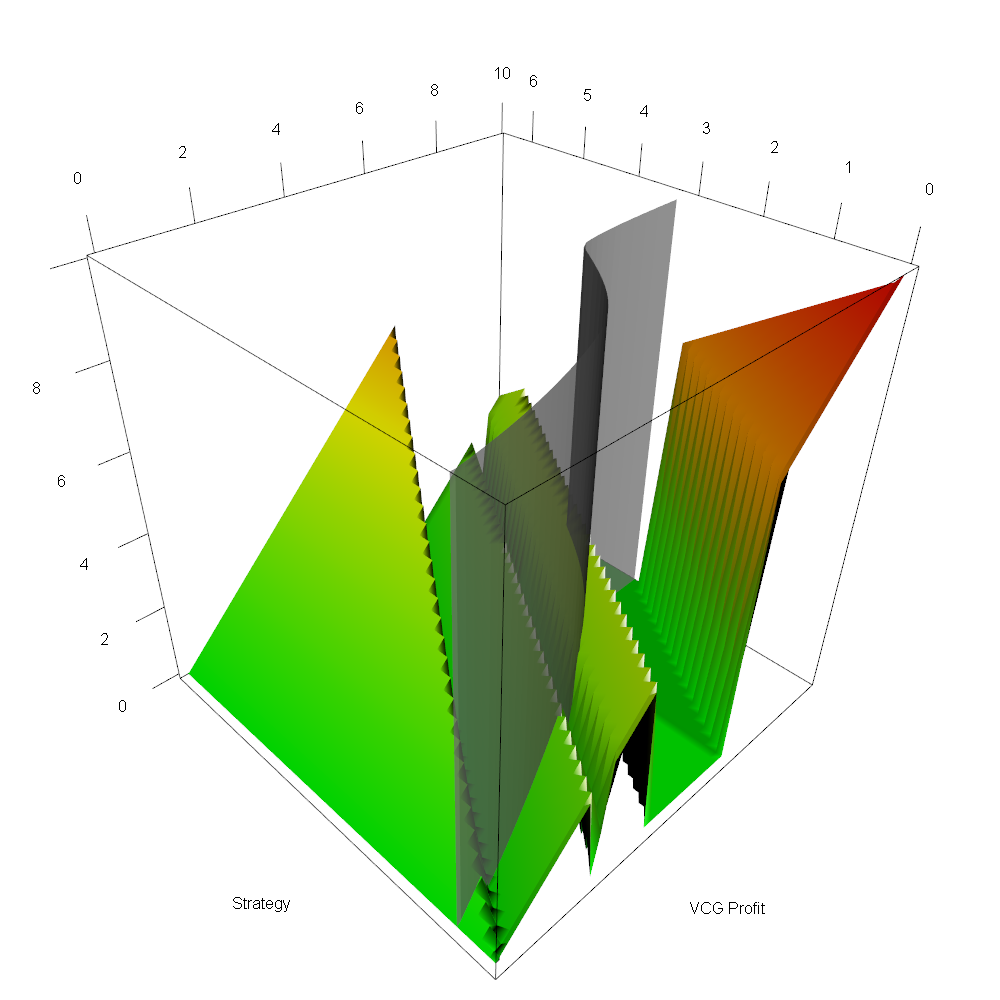}
    \label{fig:FunctionParetoStdDev2}
  }
  \subfigure[$\mu\sim$ Normal(0,5)]{
\vspace{-1em}
    \includegraphics[width=.45\textwidth]{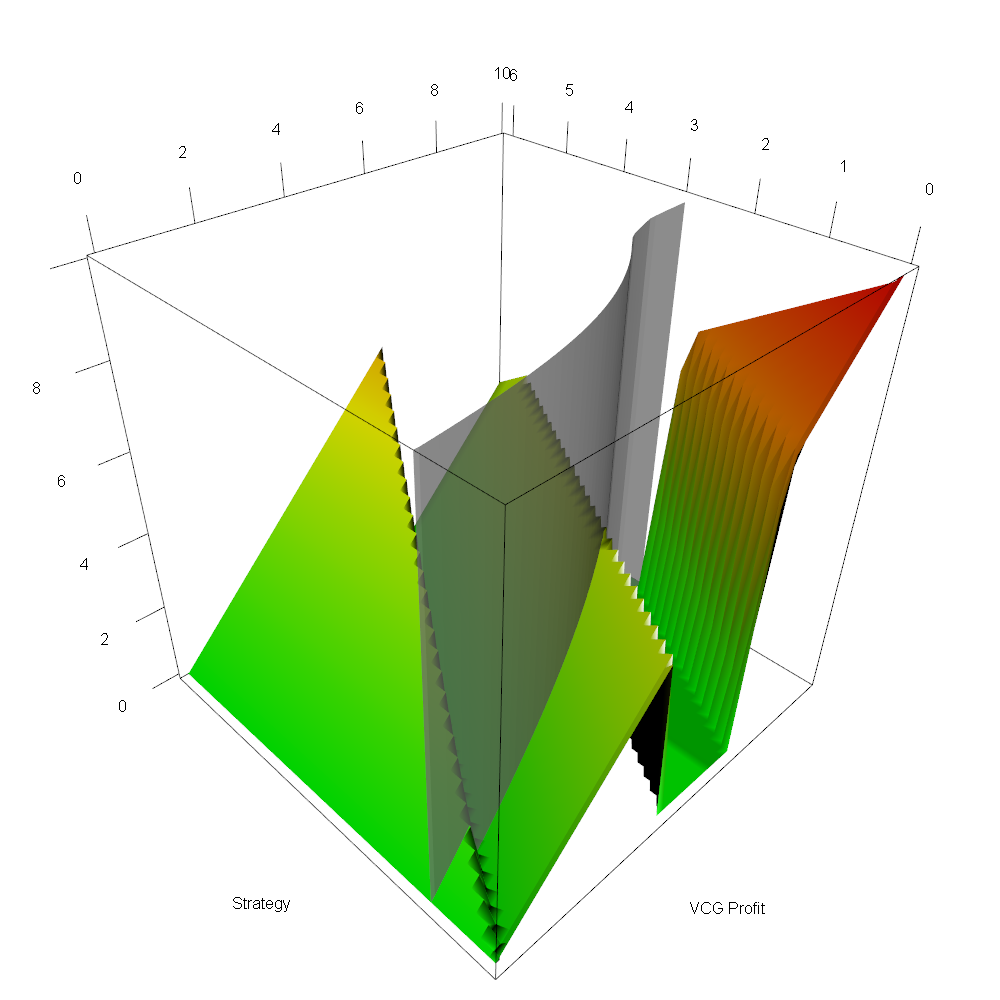}
    \label{fig:FunctionParetoStdDev5}
  }
  \subfigure[$\mu\sim$ Normal(0,10)]{
\vspace{-1em}
    \includegraphics[width=.45\textwidth]{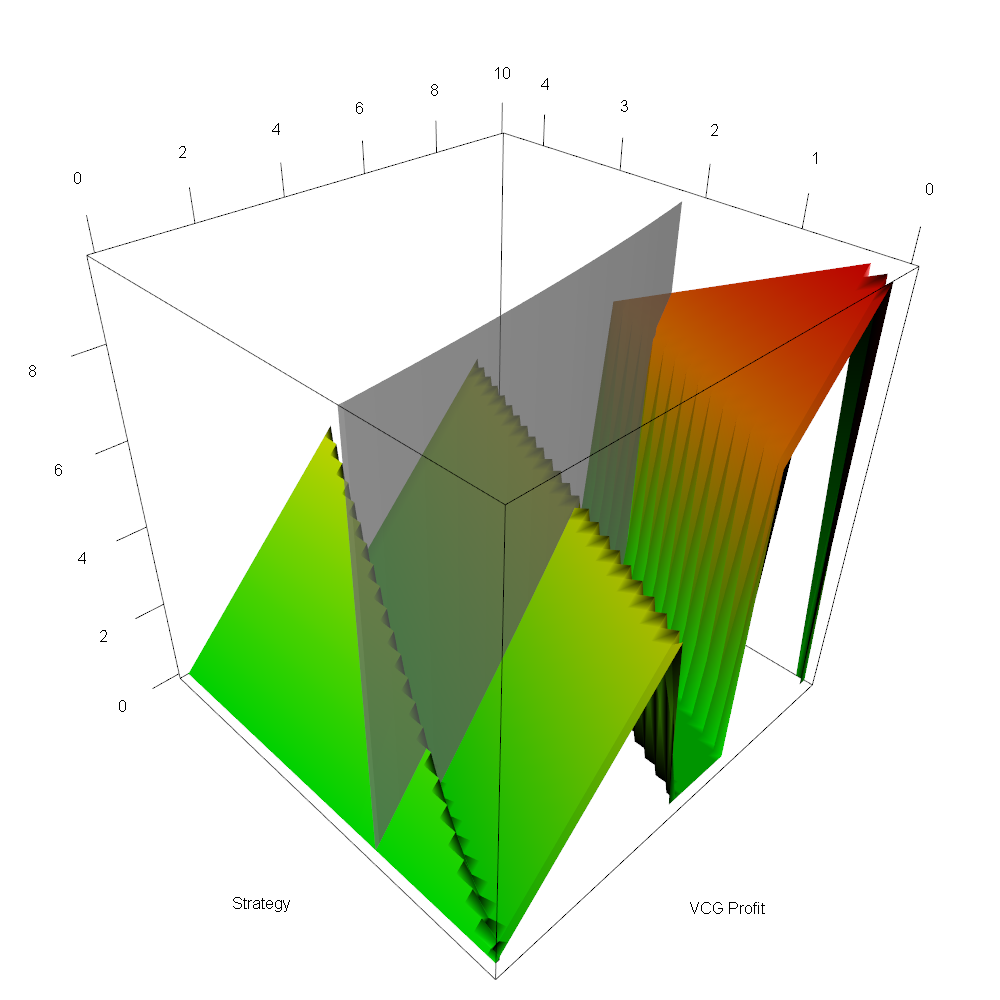}
    \label{fig:FunctionParetoStdDev10}
  }
  \subfigure[$\mu\sim$ Normal(0,1000)]{
\vspace{-1em}
    \includegraphics[width=.45\textwidth]{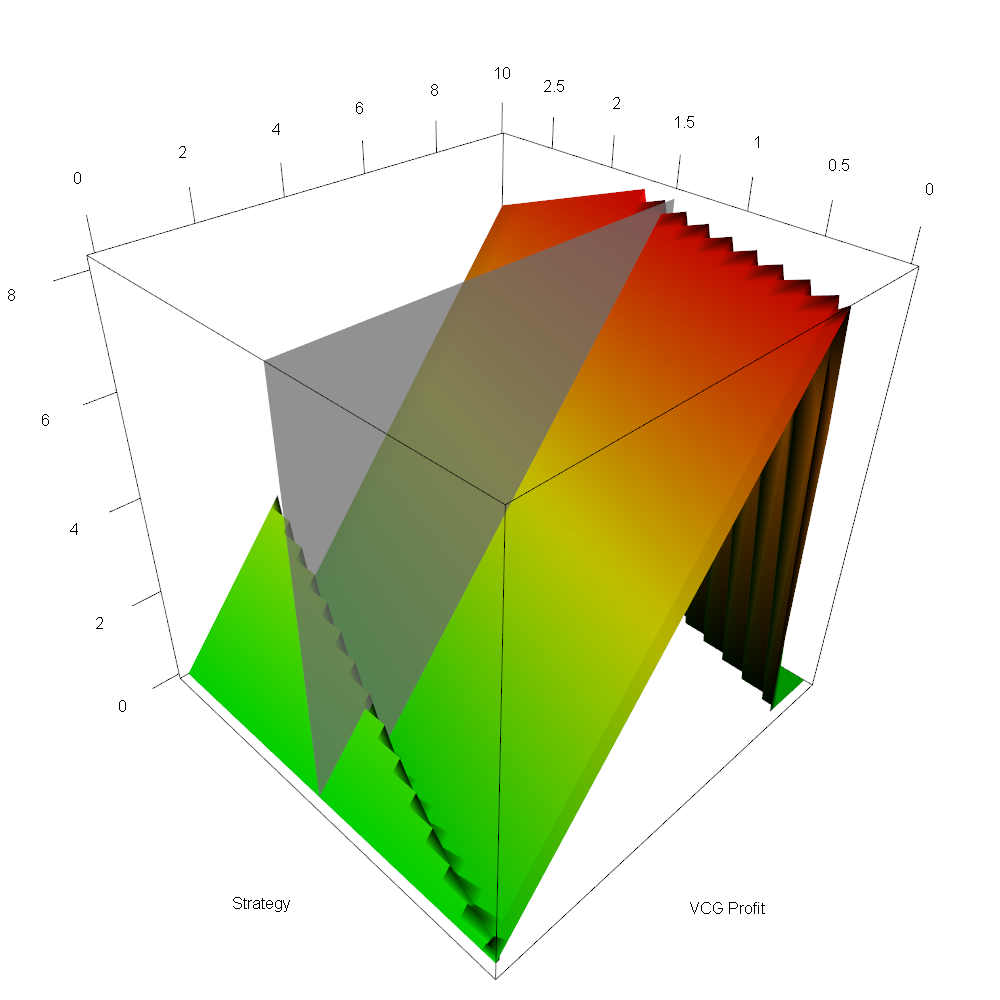}
    \label{fig:FunctionParetoStdDev1000}
  }
  \caption{Optimal \emph{Blinded Regret} payments above the critical
    value as function of both VCG potential profit and bidder
    strategy.  Here $f\sim$GPD(0,1,1) and $\mu\sim$Normal, as
    specified.  The optimal strategy is shown as the grey vertical
    surface. 
  \label{fig:FunctionParetoStdDev}}
\vspace{-1em}
\end{figure}

Finally, we turn our attention to our more complex formalism described
in sections~\ref{sec:blindedRegret} and \ref{sec:funcStratTheory}.
Figure~\ref{fig:FunctionParetoStdDev} illustrates an example from this
setup with $f$ $\sim$ GPD(0,1,1) and $\mu$ drawn from four different
Normal Distributions.  These 3-D plots show the payment above the
critical value that a bidder faces both as a function of $\psi$, or
potential profit (right axis), and as a function of potential shade he
might make at that value $\psi$ (left axis).  The figures make clear
that the 1-D payment function offsets back and to the right, as the
agent increases his shade.

We hasten to note that the pyramid region on the left is not
technically part of the payment rule itself.  Rather, in that region,
the bidder has shaded more than his VCG potential profit, and has thus
issued a bid below the critical value.  Consequently, he is now
losing, and incurring a regret equal to his full potential profit.
The region rises as it goes to the right and back, since bidders with
higher values have potentially more to lose.

Cutting through each of the figures, you can see a vertical surface.
This is a rendering of the optimal strategy taken by the bidder (under
the information set he has available), when confonted with this rule.
One can see that the amount of shaving generally increases with the
VCG potential profit that an agent has, but that the surface is not at 45
degrees, since the bidder only has partial information about his value
of $\psi$, as enforced by the \emph{blinding distribution}, $\mu$.
Interestingly, in the Figures~\ref{fig:FunctionParetoStdDev2} and
\ref{fig:FunctionParetoStdDev5}, we can see the optimal strategy
``hook'' around the mass of additional payment at the far right of the
plot.  For $\mu$ with large standard deviations, such as in
Figures~\ref{fig:FunctionParetoStdDev10} and
\ref{fig:FunctionParetoStdDev1000}, the bidder doesn't have enough
information about his setting to be able to do this.
Moreover, Figure~\ref{fig:FunctionParetoStdDev1000} is included to
illustrate that as the $\mu$ approaches the Uniform Distribution, the
agent has effectively only \emph{ex-ante} information, and thus
chooses a constant shade (illustrating how the \emph{blinded regret}
formalism generalizes the \emph{ex ante} approach.

\begin{figure}[tb]
  \centering
  \includegraphics[width=.48\textwidth]
  {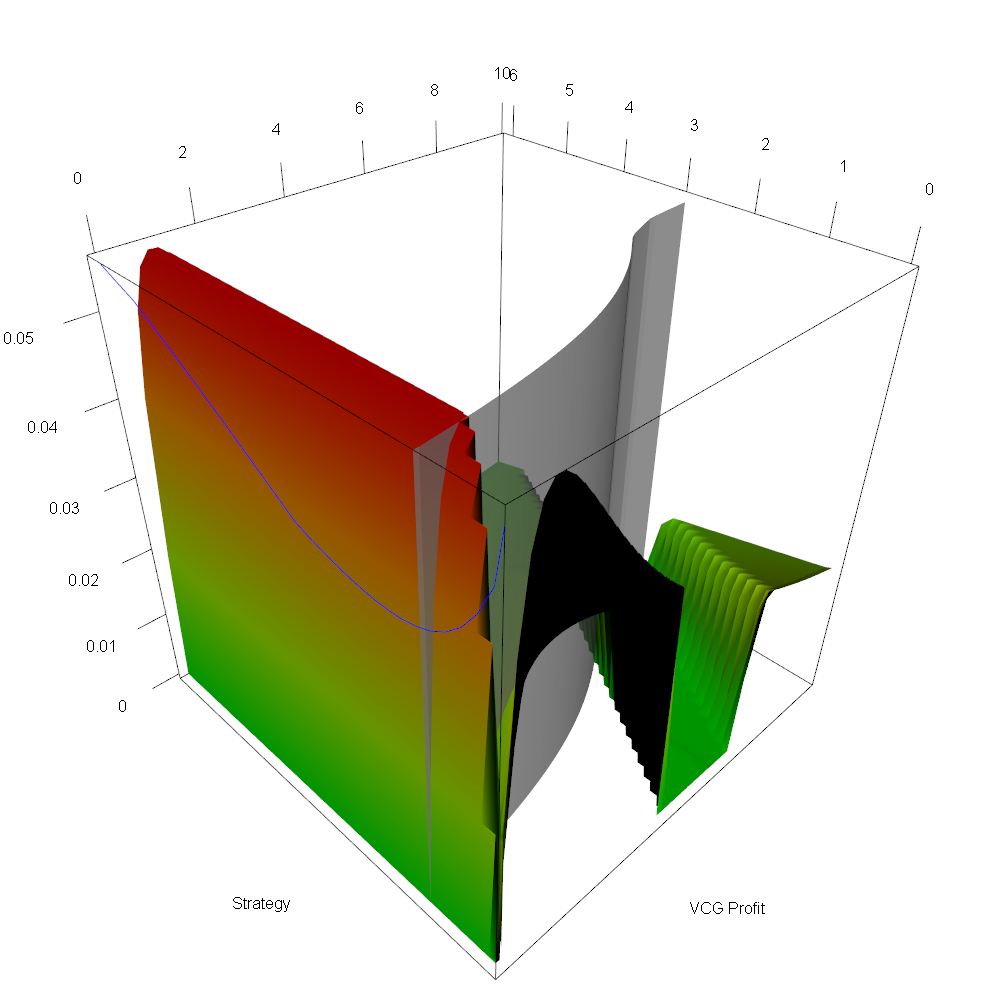}
  \caption{Optimal \emph{Blinded Regret} payments above the critical
    value scaled by the $f$ distribution and plotted as a function of
    both VCG potential profit and bidder strategy.  Here
    $f\sim$GPD(0,1,1) and $\mu\sim$ Normal(0,5).  The optimal strategy
    is shown as the grey vertical surface.  The mean regret for a
    \emph{constant} strategy is plotted on the left in blue, and is
    similar, though certainly not identical to the curve.}
  \label{fig:FunctionParetoStdDev3DPlotMultPDFStdDev5}
\end{figure}

The small dip that can be seen in the rule obtained in
figure~\ref{fig:FunctionParetoStdDev2} is not an artifact, but a
feature of the equilibrium of high-information set games (e.g. those
with $\mu$ distributions with low standard deviation).
The more information the bidder has, the more effective he is at
contorting his bid, and the more complex the center's chosen payment
rule must be in order to attempt to thwart this strategizing.
This illustrates that approximate incentive schemes under even
simplified Bayesian conditions can still be highly complex for even
moderately-informed bidders.
This complexity stems from bidders understanding their strategic
environment for exploitation.
This has implications for market design in general: we want bidders to
be able to easily calculate their valuation function, and to be able
to bid based on it -- otherwise efficiency suffers.  But, we do
\emph{not} want bidders to know their VCG potential profit $\psi$,
because knowledge of this informs them of their critical point, which
in turn enables them to strategize.  Concretely, this means agents
must be kept in the dark about the values of their competitors, which
may be hard in small markets where agents can reason directly about
individual competitors.

Figure~\ref{fig:FunctionParetoStdDev3DPlotMultPDFStdDev5} shows the
same example as Figure~\ref{fig:FunctionParetoStdDev5}, but where the
payments have been scaled by the $f$ distribution.  Although slightly
more difficult to read than Figure~\ref{fig:FunctionParetoStdDev},
this approach lets you see precisely the topology that drives the
decisions made by both the bidder and the center.  The figure
illustrates the fundamental tradeoff faced by the optimizers for a
falling distribution like GPD: small values of $\psi$ occur
frequently, so even though they are small they are important.  Large
values of $\psi$ occur rarely, but their sheer size can make them
important too.  Ultimately, both sides of the game must pay attention
to the full range in equilibrium.

\vspace{-1em}

\section{Discussion}


\vspace{-1em}

Although there are numerous domains where combinatorial exchanges
could provide tangible economic benefits, including spectrum
reallocation and computational resource allocation, a significant
obstacle has prevented them from being used in practice: a reasonable
payment rule with good incentive properties.

In this work, we have presented the first comprehensive method for
constructing such a rule.  To do so, we have employed a construction
that reduces the complexity of the payment rule design problem to a
single dimension, by appealing to the same principles that enable the
VCG mechanism to be strategyproof.  In particular, the rule is defined
as a single dimensional function, whose argument is the amount in
excess of the critical value the participant has bid, and whose result
is the amount in excess of the critical value that he is then asked to
pay.  This amount is always exactly zero for VCG, but for our rules it
is a positive number for at least some bidders such that in total the
budget balance constraint can be met, while simultaneously attaining
our definition of approximate incentive compatibility.  In this sense,
it reduces to a payment rule that directly reasons about the profit
that agents obtain for their allocated bundle, without worrying about
the combinatorial complexities that lead to that bundle being chosen

In conjunction with this, we have also constructed a corresponding
model for bidder behavior, again reduced to the single dimension of
bidder profit.  With this reduction, we are in a position to offer a
novel definition of approximate incentive compatibility, where we
consider the increased profit an agent might attain by optimally
misreporting based on a distribution of the available potential VCG
profit in the domain, when the other agents are playing their
Bayes-Nash equilibrium strategies.  Rather than assuming agents have
direct access to this distribution, which would result in an
\emph{ex-post} information set, we instead introduce a ``blinding''
distribution, which represents an agent's bounded knowledge about its
potential profitability in the domain.  By compounding the blinding
distribution with the potential VCG profit distribution, we obtain our
model of agents' fuzzy knowledge of their domain.  This model has the
advantage of the tractability of a single-dimensional \emph{ex-post}
analysis, while retaining much of the economic structure of a far more
desirable \emph{ex-interim} model.

By combining this model for the center and for the agents together, we
can construct the full mechanism.  This mechanism operates exactly
like VCG, except that it calculates different prices.  These prices
are based on particular distributional information about how agents
bid in equilibrium, and about a given agent's true value relative to
its critical value (i.e. its true VCG discount).  
%
%
%
A critical contribution of this framing of the pricing problem is that
we cast the problem as a
meta-game between the center and the agents playing the game.
We argue that this structure is a general property of problems in
\emph{mechanism design}: the mechanism and the participants are at
odds with each other, each trying to box in the other in order to
attain its own goals.
%

To obtain concrete rules in our paradigm in the general case, it is
necessary to go beyond closed-form solutions and employ numerical
methods to solve for the equilibrium of the meta-game.  Such solutions
simultaneously offer approximately IC payment rules under our new
blinded regret information set, and the optimal agent strategies when
agents face these rules.  We offered such a numerical method, adapted
from the traditional damped interated best-response algorithm for
solving BNEs.
We then tested the approach on several distribution classes and a wide
range of parametrizations, including those that have previously been
shown to be a good fit with real data.  Using these distributions we
were able to generate the Large and Small rules computationally, in
precisely the settings where they have been shown effective in
practice.  Moreover, going beyond these settings, our method produces
a wide variety of novel, yet easily implementable, payment rules.

\pagebreak
\bibliographystyle{plainnat} 
\bibliography{MetaGamePricing}

\pagebreak
\appendix
\section*{Appendices}
\renewcommand{\thesubsection}{\Alph{subsection}}
\setcounter{section}{0}

\section{Existing formulations of approximate incentive compatibility}
\label{app:existingApproxIC}

The most widely used approximate incentive compatibility concept is
the \emph{Worst Case Ex-Post Regret}:
\begin{equation*}
\text{WCRegret}_i = \max_{\mathbf{v}_{\sni}} \max_{v_i} \max_{\hat{v}_i} 
[\pi_i(v_i,\hat{v}_i, \mathbf{v}_{\sni}) - \pi_i(\mathbf{c})]
\end{equation*}
This measures across all possible agent values and reports the most
that an agent might gain by misreporting, given what he knows after
the mechanism is complete.  In other words, it is the ``Full
Information'' setting, and thus reduces our BNE to a regular Nash
equilibrium.
Importantly, though, it will not minimize efficiency loss.  Efficiency
is a global criterion, seeking to match total value (or equivalently
expected value per participant) of the reported and true allocations.
WCRegret, by contrast seeks to minimize the worst case difference in
reported versus true allocative value for individual participants,
$v_i(\lambda_i^\ast) - v_i(\hat{\lambda}_i)$.  A given mechanism may
exhibit a large such loss on certain participants in order to reduce
the overall loss in aggregate, thereby improving efficiency.
Yet, the measure has the virtue of being both understandable and easy
to calculate.  The definition works well when it can be expected to be
small in magnitude, hence its other name ($\epsilon-$strategyproofness
-- \citeauthor{roberts1976incentives},
\citeyear{roberts1976incentives}).  The current state-of-the-art
rule, Threshold \citep{Parkes2001threshold}, minimizes this value,
e.g.  $L_\infty(\Delta, \Delta^{VCG})$.
However, when, as in the case of CEs, we expect the \emph{ex-post}
regret to be large, the measure is overly conservative: it tells us
solely about a very rare and very massive worst case -- preventing us
from targeting the typical cases for design.  High efficiency can
require accepting significant value loss to a tiny minority of
individual participants, who incur exceedingly poor WCRegret.

Consequently, we might instead an expectation over participant values
instead, yielding expected \emph{ex-post} regret:
\begin{equation*}
\text{ECRegret}_i = \Expect_{\mathbf{v}_{\sni}} \Expect_{v_i} \max_{\hat{v}_i} 
[\pi_i(v_i,\hat{v}_i, \mathbf{v}_{\sni}) - \pi_i(\mathbf{v})]
\end{equation*}
Beause it is still a full information analysis, this retains appealing
computational properties.  However, it consequently also retains an
unrealistic information set, requiring us to assume that participants
have full information about each other which is unrealistic in most
settings.

To improve the information assumptions, we instead may wish to
consider \emph{Ex-Ante Deviation Incentive} (ADI):
\begin{equation*}
 \text{ADI}_i = \max_{\hat{v}_i}
   \left[ \Expect_{v_i}
     \left[ 
        \Expect_{\mathbf{v}_{\sni}} [\pi_i(v_i,\hat{v}_i,\mathbf{v}_{\sni}) -
                                \pi_i(\mathbf{v})]
     \right]
   \right]
\end{equation*}
In this formulation agents have only probabilistic information about
the value profile over all agents, including their own values.  Thus
the information set is problematic for the opposite reason from an
\emph{ex-post} analysis: we are now assuming agents know too little,
rather than too much.  But, because the expection over value of the
other agents is taken relative to their true values, not relative to
their reports, formulating the distribution does not require
repeatedly solving for the BNE outcome.  Consequently, ADI is
relatively computationally tractable, even though it typically will
require solving a winner determination problem within the stochastic
optimization.  We therefore consider a pricing model based on this
model, when adapted to our methodology, in
section~\ref{sec:constStratTheory}.

What we really want, though, is an information set, where agents know
their own values exactly but the values of the other agents only in
expectation.  This yields the \emph{Ex-Interim Deviation Incentive}
(IDI):
\begin{equation*}
 \text{IDI}_i = \Expect_{v_i}
   \left[ \max_{\hat{v}_i}
     \left[ 
        \Expect_{\mathbf{v}_{\sni}} [\pi_i(v_i,\hat{v}_i,\mathbf{v}_{\sni}) -
                                [\pi_i(\mathbf{v})]
     \right]
   \right]
\end{equation*}
This represents a tremendous improvement in fidelity: if the
expectation over the values of others in the first term is instead
taken relative to BNE reports, then this would measure the value of
the expected best-response and directly calculates our gold-standard
measure, trending to $0$ as efficiency goes to $1$.\footnote{%
$
\Expect_{v_i}
   \left[ \max_{\hat{v}_i}
     \left[ 
        \Expect_{\hat{\mathbf{v}}_{\sni}} [\pi_i(v_i,\hat{v}_i,\hat{\mathbf{v}}_{\sni})] -
        \Expect_{\mathbf{v}_{\sni}} [\pi_i(\mathbf{v})]
     \right]
   \right]
$
}%
Despite this, typically the distribution is still taken relative to
the truthful reports, to match the definition of \emph{ex-ante}, while
adding only to the information set participants' knowledge of their
own value.  Regardless of which distribution is considered, though,
IDI is typically computationally even more challenging than
\emph{ex-ante} since the maximization is now embeded within a larger
expectation.

\pagebreak
\section{Proof of observations}
\label{app:obs}

\begin{proof}[Observation~\ref{ob:PSI}]
  (1) Follows directly from the definition of the \emph{critical
    value}, namely that when $\hat{v}_i < v_i^C$, then by definition
  $\hat{V}^\ast(\hat{v}) <
  \hat{V}_{\sni}^\ast(\hat{\mathbf{v}}_{\sni})$ and the marginal
  economy without $i$ has greater value than the main economy and
  $\lambda_{\sni}^\ast$ will be chosen over $\lambda^\ast$.  (2) Is
  immediate.
\end{proof}
\begin{proof}[Observation~\ref{ob:PSI}]
  Follows from the definitions of $\pi_i, p_i$, and
  $\Delta_i^{VCG}$ and from VCG being strategyproof. 
\end{proof}
\begin{proof}[Observation~\ref{ob:outEq}]
  Is equivalent to the foundational property of VCG that for reports
  above $v^c_i$ (truthful or otherwise) $i$'s profit under VCG is his
  marginal impact on the economy, a constant independent of the
  reported value $\hat{v}_i \geq v^c_i$,
\end{proof}

\section{Full computational mechanism design problem}
\label{app:AMD}

Without the simplifications we employ in this papaer, the full
Automated Mechanism Design \citep{conitzer2004amd} approach to the
pricing problem, would be to solve:
\begin{align}
  \argmin_{\mathbf{p} \in \{ \prod_N \mathcal{P}(M) \times
  \left(\prod_N \mathcal{P}(M) \rightarrow \prod_N \mathbb{R} \right)
  \} } \quad & \max_{i \in N} \text{ApproxSPGoal}_i
  \label{eq:amdApproach} \\
  \text{s.t.}\quad\quad\quad\quad\quad\quad\quad & \,\,\text{IR, Weak
  BB}\notag
\end{align}
where $\text{ApproxSPGoal}_i$ is any of the deviation incentive
measures from appendix~\ref{app:existingApproxIC}, or the new one
definind in equation~\ref{eq:blindedRegretDI}.
Unfortunately though, the variable in this program is a vector-valued
(one entry per bidder) \textit{payment function} and the objective
functional is a stochastic optimization containing embedded winner
determination problems.  Thus, the program is intractable for all but
the most trivial of instances.\footnote{See
  \citep{lubin2010combinatorial} for a simple example solved in a
  discretization of this formulation.}  Consequently, we seek to gain
traction by a suitable simplification that will enable reasonable
computation.

\pagebreak
\section{Ratio method for solving the center problem}
\label{app:ratio}

Here we present an alternative way to solve the problem shown in
Section~\ref{sec:methodCenter}.

Consider the following change of variables in the
original continuous BB constraint from \eqref{eq:constCenter}: $\psi
\rightarrow \psi-s$:
\begin{equation}
\int_s^{U+s}  r(\psi)f(\psi-s)d\psi\geq k
\end{equation}
Now consider a starting choice for the $r$ function $r(\cdot)=0$.  We
then wish to modify $r$ into $r(\rho) = \epsilon$ for some $\rho$ that
will optimally progress towards satisfying the BB constraint while
simultaneously doing the least damage to the objective function.  To
do so, we should choose
\begin{equation}
  \rho = \argmax_{\psi} \frac{f(\psi-s)}{f(\psi)}
\end{equation}
because it is this ratio that determines where progress is best made
trading off adding to the constraint and to the objective.

\begin{figure}[b]
\centering
  \centering
  \includegraphics[width=.48\textwidth]{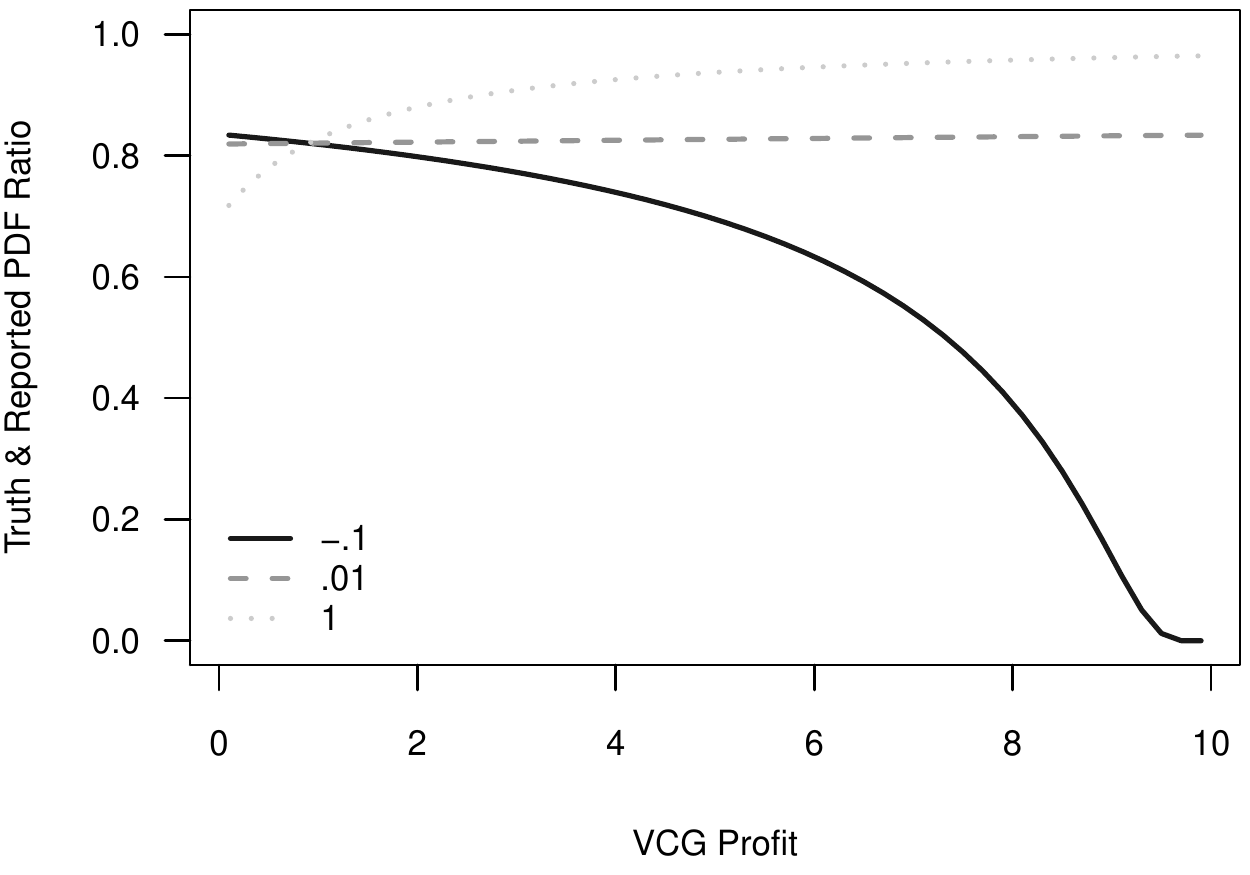}
  \caption{\emph{Ex-ante} pdf ratios plotted for several examples of
    the Pareto shape parameter}
  \label{fig:ConstantParetoShapeRatios}
\end{figure}

This insight offers an alternative method to solve for $r$,
provided in algorithm~\ref{algo:ratio}.
We can also plot this ratio to gain insight into what is driving
particular rule choices, as in
Figure~\ref{fig:ConstantParetoShapeRatios}.  Here we see a plot of the
ratios, one line for each of the distributions being considered.  The
plot shows a very different shape for each of the parametrizations.
For $-.1$, a finite distribution, the ratio slopes downwards and
consequently the corresponding rule shown in
Figure~\ref{fig:constantParetoShapeRegret} asks for payments from only
the bidders with small $\psi$.  The rule selected in this case is the
Large rule, which gives all the available surplus to bidders with
large VCG potential profits.
As we described in section~\ref{sec:distribution}, all possible rules
consistent with the boundary and budget constraints are valid for the
Exponential Distribution under the basic \emph{ex-ante} formulation.
By picking a shape parameter just larger then zero we can see the
limiting case of the GPD as it approaches the exponential
distribution.  In this case the ratio is nearly flat (it would be
perfectly flat for the Exponential Distribution).
When the scale parameter is $1$, a heavy-tailed distribution, the
ratio is rising.  This leads to a rule that asks only those bidders
with the largest $\psi$ to pay.  Accordingly, this parametrization
selects the Small rule, which gives all the available surplus
to bidders with the smallest VCG potential profits.

\begin{algorithm}
\DontPrintSemicolon 
\KwIn{A distribution $f$ (or $g$), and strategy $s \in \mathbb{R}$ and 
      required regret $k$}
\KwOut{A discretized regret function $r$}
Create $B$ discrete bins\;
$\rho_b \gets \frac{1}{\overline{b}-\underline{b}} 
                       \int_{\underline{b}}^{\overline{b}} \frac{f(x)}{f(x-s)}dx$\;
\For{$b \in B $ in descending order of $\rho_b$} {
  $z \gets \frac{k}{F(\overline{b})-F(\underline{b})}$\;
  $r_b \gets \min($  \sout{$b$}  $, z)$\;
  $k \gets k-r_b z$\;
  \If{$k \leq 0$}{ 
    break\;
  }
}
\Return{$r$}
\caption{Find an approximate regret function by the ratio method}
\label{algo:ratio}
\end{algorithm}

\pagebreak
\section{$f$ $\sim$ Burr XII. a further \emph{Ex-Ante} result }
\label{app:exanteBurr}

\begin{figure}[bh]
  \centering
  \includegraphics[width=.48\textwidth]
{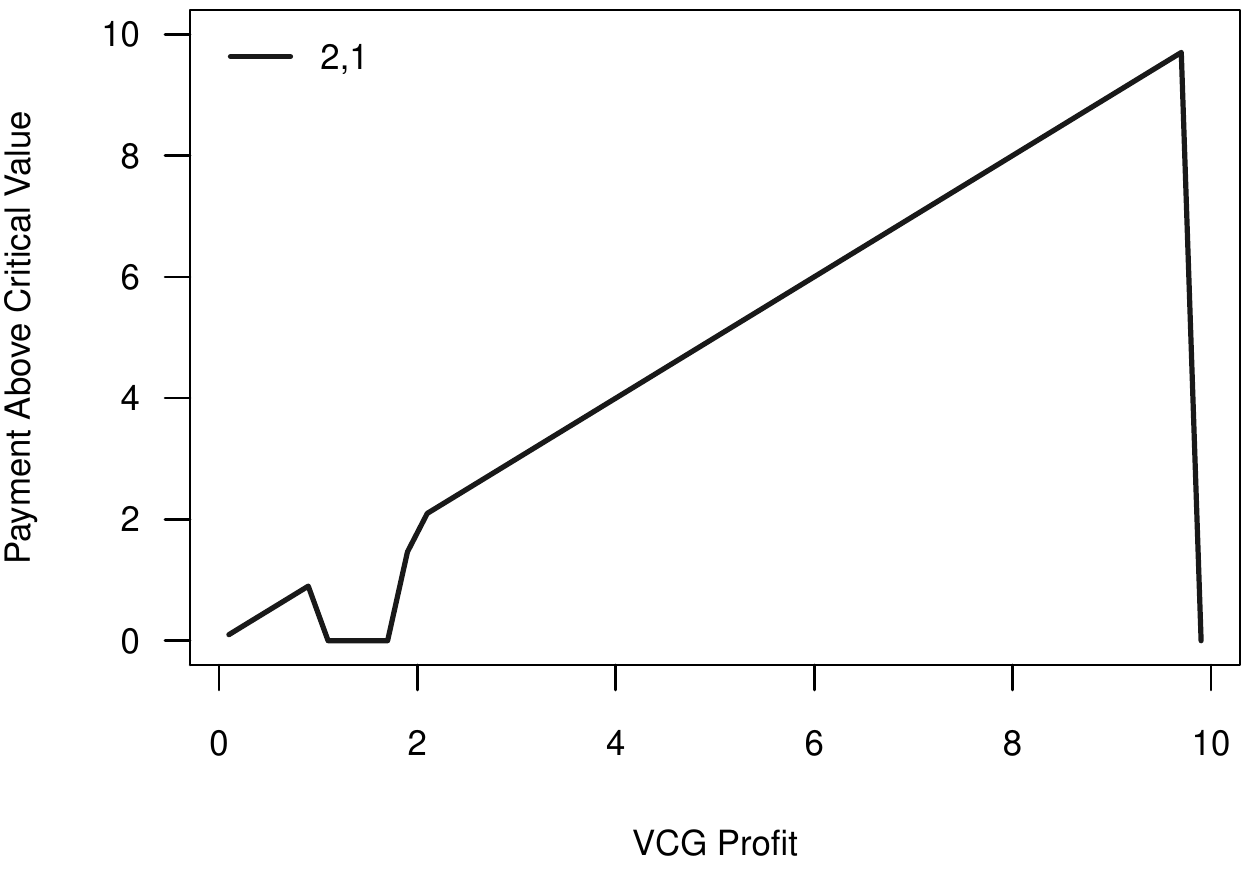}
  \caption{\emph{Ex-ante} payments above the critical value with a 
           Burr XII distribution parameterized as $(2,1)$}
  \label{fig:ConstantBurrShapeRuleRegret21}
\end{figure}

We have also run a series of experiments using a generalization of the
Pareto Distribution, known as the Burr XII Distribution, which has two
shape-parameters.  When the first parameter is $1$, the Burr XII
Distribution is the same as a Pareto Distribution.  When the second
parameter is $1$, we have a Log-Logistic Distribution
\citep{burr1942cumulative}.
Figure~\ref{fig:ConstantBurrShapeRuleRegret21} shows the $r$ function
obtained from a parametrization of $(2,1)$, which produces a peaked
distribution in the family of Log-Logistic functions.  Here we see
that the bidders near the peak receive their full VCG potential profit
and obtain all the surplus.  The bidders with either small or large
VCG potential profits are given no discount at all.  Essentially, this
rule will give all of its surplus to a subset of values in the middle.
This example shows how dependent the outcome of the optimal payment
rule is on the shape of the distribution.  Nonetheless, the optimal
strategy has not changed, and remains a shade of $0.2$.

\end{document}